# Conformal Metamaterials with Active Tunability and Self-adaptivity for Magnetic Resonance Imaging

Ke Wu[1,3], Xia Zhu[1,3], Xiaoguang Zhao[2,3], Stephan W. Anderson[2,3*], and Xin Zhang[1,3*]

[1*]Department of Mechanical Engineering, Boston University, Boston, MA 02215, United States.

[2*]Department of Radiology, Boston University Medical Campus, Boston, MA, 02118, United States.

[3]Photonics Center, Boston University, Boston, MA 02215, USA.

*Corresponding authors. E-mail(s): xinz@bu.edu; sande@bu.edu;

Contributing authors: wk0305ok@bu.edu; xiaz@bu.edu; zhaoxg@bu.edu;

**Abstract**

Ongoing effort has been devoted to applying metamaterials to boost the imaging performance of magnetic resonance imaging (MRI) owing to their unique capacity for electromagnetic field confinement and enhancement. However, there are still major obstacles to widespread clinical adoption of conventional metamaterials due to several notable restrictions, namely: their typically bulky and rigid structures, deviations in their optimal resonance frequency, and their inevitable interference with the transmission RF field in MRI. Herein, we address these restrictions and report a conformal, smart metamaterial, which may not only be readily tuned to achieve the desired, precise frequency match with MRI by a controlling circuit, but is also capable of selectively amplifying the magnetic field during the RF reception phase by sensing the excitation signal strength passively, thereby remaining 'off' during the RF transmission phase and thereby ensuring its optimal performance when applied to MRI as an additive technology. By addressing a host of current technological challenges, the metamaterial presented herein paves the way toward the wide-ranging utilization of metamaterials in clinical MRI, thereby translating this promising technology to the MRI bedside.

**Introduction:**

Metamaterials, rationally designed assemblies of subwavelength unit cells (meta-atoms), enable the unique capacity to tailor the effective properties of artificial materials, thereby realizing characteristics not found in naturally occurring materials, such as left-handed materials (LHM) or negative refractive index materials (NIM)[1-3]. With extraordinary electromagnetic (EM) properties, myriad efforts in developing metamaterials have not been limited simply to demonstrating that the metamaterials' properties extend beyond those of natural materials by breaking generalized limitations of refraction and reflection, but have also been focused on developing novel metamaterial-enabled technologies to facilitate a range of practical applications from the microwave to the optical regime, such as cloaking devices[4], perfect absorbers[5], super lenses[6], and metamaterial antennas[7], among others. One notable property of metamaterials is the near-field enhancement due to their ability to confine the energy of incident radiation to a sub-wavelength region in the vicinity of the metamaterials. On resonance, such metamaterials generate an intense and localized electric field at the edges of their narrow capacitive gaps where the induced transient charge accumulates. Such electric field confinement and its corresponding enhancement due to metamaterials has enabled a wide range of applications, including resonant optical antennas[8], plasmon optical tweezers[9], and nonlinear-based devices for phase conjugation[10], to name a few. As in the case of electric field confinement, a similar confinement effect also occurs to the magnetic field. When metamaterials are excited by an incident wave at their resonance frequency, an induced circulating current distributes along their conductive or metallic structures, leading to a resonant enhancement of the local magnetic field near the regions of the peak transient current. This capacity of magnetic metamaterials for magnetic field enhancement has enabled their application to wireless power transfer[11,12], high-sensitivity sensing[13], and second-harmonic generation[14], among others. Leveraging this unique magnetic field enhancement property, a variety of metamaterials and metasurfaces, composed of diverse arrays of unit cells of various configurations such as conducting "Swiss rolls", parallel metallic wires, or helical coils, have been utilized in magnetic resonance imaging (MRI) to enhance the signal-to-noise ratio (SNR) by boosting the radiofrequency (RF) magnetic field strength[15-17]. However, there are substantial barriers to overcome in order to achieve the eventual clinical adoption of metamaterials in MRI, including their typically rigid and bulky structures, a susceptibility to variations in their resonance frequency, and an inherent linearity exhibited by these conventional classes of metamaterials designed for use in MRI.

The majority of the reported metamaterials for enhancing MRI systems are typically constructed using bulky and rigid structures, which not only compromises patient comfort, but also and importantly, markedly limits the optimized imaging of curved surfaces, such as the brain, neck, or musculoskeletal system (knee, ankle, etc.), as the SNR gains of metamaterials decays rapidly as a function of distance from the metamaterial surface. More importantly, the planar configured metamaterials are only sensitive to linearly polarized magnetic fields, while the MRI signal is a circularly polarized field[18,19], therefore, only half of the RF reception signal is enhanced by conventional planar metamaterials. Furthermore, to ensure the optimal performance of metamaterials in MRI, a precise match in working frequency between the metamaterial and MRI system should be satisfied in order to achieve an optimal resonant magnetic field enhancement. However, the resonance frequency of metamaterials is susceptible to their local environments and the presence of materials with different permittivities, thus, the metamaterial resonance frequency may shift to undesired values when in proximity to patients of varying body composition (differing degrees of water, fat, muscle, or bone) during an MRI scan. Of note, as is the case for the conformal metamaterial reported herein, the structural deformation may also exert an influence on its resonance frequency by altering the coupling coefficient between unit cells in the metamaterial. As a result, the capacity for frequency tunability is highly desired in a metamaterial in order to ensure an optimal frequency in varying local environments and different configurations. Metamaterials incorporating active materials as components of their constitutive elements or integrating structures enabling physical perturbation have been reported to yield the capacity for EM tunability[20-24]. In terms of metamaterials for MRI, the focus herein, we previously reported a class of mechanically tunable metamaterials inspired by auxetics to manipulate the interactions between unit cells, thereby tuning the resonance frequency by varying the density of the meta-atom array[25]. Though simple and straightforward to implement, mechanically tunable metamaterials are practically challenging to embed into the MRI system to form an efficient feedback loop in order to achieve automated frequency matching. Beyond the challenge of frequency matching, another notable limitation of conventional metamaterials applied to MRI is their intrinsic linearity, leading to a magnetic field enhancement regardless of RF power, thereby resulting in field enhancement during both RF reception and transmission phases. While the enhancement of the RF field $B_1^-$ during the reception phase results in a desired boost of SNR, the amplification of the RF field $B_1^+$ during the transmission phase leads to undesired problems, such as unpredictable deviations in flip angle (FA), suboptimal performance, and potential safety concerns related to an increase in the specific absorption rate (SAR). Nonlinear metamaterials, whose EM behavior is not only dependent on frequency but also affected by the intensity of the incident electromagnetic field[26-31], provide a promising method to design self-adaptive metamaterials for MRI, resulting in an 'off' state during RF transmission and a desired 'on' state during RF reception. Herein, through the development and validation of a conformal metamaterial featuring active tunability and self-adaptivity, we intend to address the aforementioned ongoing limitations in conventional metamaterials, easing the ongoing translation of these exciting technologies towards clinical use.

## 2. Results
## 2.1 Modeling of the meta-atom and metamaterial characterization

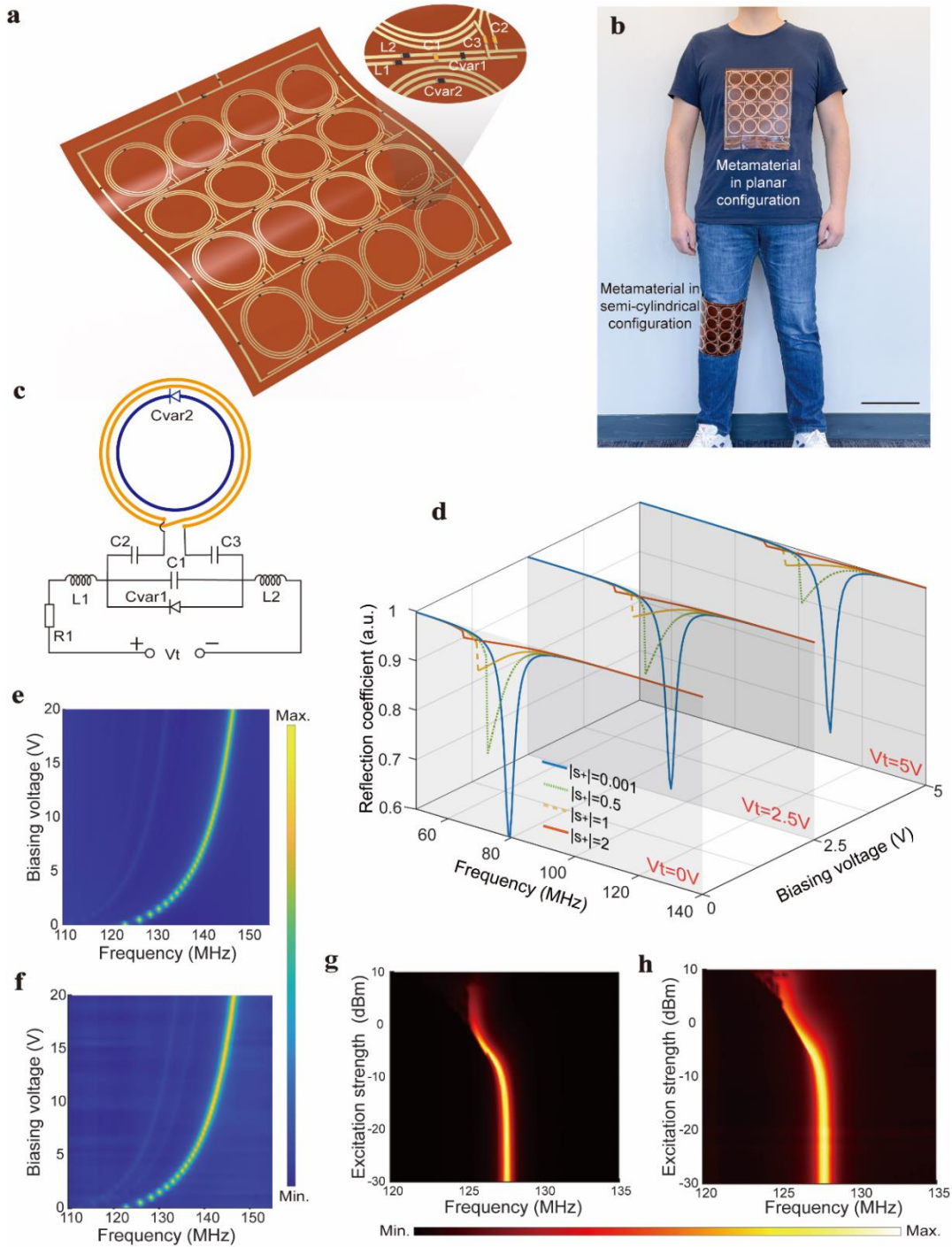

**Fig. 1 | Concept and characterizations of the metamaterial. a,** Image of the metamaterial in a flexible state. Inset: Magnified view of the elements loaded in each unit cell. **b**, Photograph of a human wearing the proposed metamaterial. Scale bar, 20 cm. **c,** Schematic diagram of the equivalent circuit of the unit cell in the metamaterial. **d,** Theoretical frequency responses of the unit cell at different excitation strengths for varied biasing voltages, e.g., Vt = 0 V, Vt = 2.5 V, and Vt = 5 V, respectively. **e, f,** Experimentally measured reflection coefficient spectrum of the metamaterial as a function of biasing voltage when metamaterial is configured in planar (**e**) or semi-cylindrical shapes (**f**). **g, h,** Experimentally measured reflection coefficient spectrum as a function of excitation strength when metamaterial is configured in planar (**g**) or semi-cylindrical shapes (**h**).

To address the aforementioned general and inherent limitations of conventional metamaterials, we developed a conformal, active tunable metamaterial with self-adaptivity for MRI applications. The ultra-thin, flexible structure serves to ensure a conformal approximation between the metamaterial and the surface of the object of interest as well as alleviating potential patient discomfort due to inherently rigid, bulky structures. The designed metamaterial is composed of unit cell array fabricated on an ultra-thin flexible polyimide substrate, as shown in Fig. 1a and a human wearing the conformal metamaterial in a 4 × 4 array fashion is shown in Fig. 1b. The unit cells feature a controlling circuit loaded spiral resonator (CCLSR) inductively coupled with a varactor loaded ring resonator (VLRR), whose equivalent circuit diagram is plotted in Fig. 1c. The dimensions of the CCLSR and the VLRR, the components employed in the circuit, as well the fabrication process, are described in detail in Methods, Supplementary Note 1, Supplementary Table 1, and Supplementary Fig. 1. In the controlling circuit, the capacitors $C_2$ and $C_3$ are used to provide the 2-turn spiral coil with an isolation from the biasing voltage, $C_1$ is employed to adjust the resonance frequency of the CCLSR to ~100MHz, and $R_1$ is inserted to limit the current in the circuit. Since the unit cells in the metamaterial are connected electrically by the biasing voltage lines, two inductors $L_1$ and $L_2$ are adopted to provide isolation from one another. The varactor $C_{var1}$ provides a variable capacitance controlled by the biasing voltage $V_t$. The resonance frequency of the CCLSR may be expressed as $\omega = 1/\sqrt{LC}$, in which the inductance L is predominately attributed to the inductance of the spiral coil, while the effective capacitance C may be expressed by:

$$C_{CCLSR} = C_s + \frac{1}{\frac{1}{C_1 + C_{var1}(V_t)} + \frac{1}{C_2} + \frac{1}{C_3}} = C_s + \frac{1}{\frac{1}{C_1 + C_0\left(1 + \frac{V_t}{V_P}\right)^{-M} + C_P} + \frac{1}{C_2} + \frac{1}{C_3}} \quad (1)$$

in which $C_s$ is the distributed capacitance in the spiral coil, $C_0$ is the initial capacitance of varactor $C_{var1}$, M is a fitting exponent, and $V_P$ is the intrinsic potential of the varactor obtained from the data sheet. $V_t$ is the voltage across the varactor applied by the controlling circuit. As for the VLRR component, its effective inductance mainly results from the inductance of the ring loop, with its capacitance being expressed by[26,31]:

$$C_{VLRR} = C_{var2}(V_D) = C_0\left(1 - \frac{V_D}{V_P}\right)^{-M} \quad (2)$$

in which $V_D$ is the voltage across the varactor $C_{var2}$. Unlike the driving voltage $V_t$ across the varactor $C_{var1}$ applied by the controlling circuit, which could be adjusted at will, the voltage across the varactor $C_{var2}$ results from the rectifying effect of the diode junction when excited by an RF wave. When excited by an RF wave with a relatively low power, the induced voltage $V_D$ is much smaller than $V_P$, resulting in an effective capacitance of the VLRR close to the initial capacitance of the varactor $C_0$. However, when excited by an RF wave of high power, the oscillation strength of the VLRR becomes stronger and induces a higher driving voltage across the varactor $C_{var2}$, which in turn increases the capacitance of the varactor, thereby shifting the resonance frequency to a lower value. As a result, this excitation power-dependent response endows the VLRR with a desired self-adaptivity to the incident excitation power strength. Of note, the integrated electronic components (capacitors, inductors, and varactors) employed in the metamaterial will introduce series resistance, which inevitably introduce loss and reduce the Q value of the metamaterial. In order to achieve the optimal performance of the metamaterial, we investigated and evaluated the impact of the series inner resistance on the SNR enhancement performance, which is described in detail in Supplementary Note 2. When these two resonators, i.e., the tunable CCLSR and the self-adaptive VLRR, are placed in close proximity to one another, they strongly interact with one another, which may be expressed by the coupling factor *k*. Their resonance response may be mathematically described using the coupled-mode-theory (CMT)[31-33]. Assuming the external excitation signal is a harmonic function with frequency ω (i.e., $s_+=|s_+|e^{j\omega t}$), the oscillating amplitude of the two resonators may be calculated by numerically solving the matrix equations:

$$j\omega \begin{bmatrix} a_1 \\ a_2 \end{bmatrix} = j \begin{bmatrix} \omega_1(V_t) + j\left(\frac{1}{\tau_{e1}} + \frac{1}{\tau_{o1}}\right) & k \\ k & \omega_2(|a_2|) + j\left(\frac{1}{\tau_{e2}} + \frac{1}{\tau_{o2}}\right) \end{bmatrix} \begin{bmatrix} a_1 \\ a_2 \end{bmatrix} + \begin{bmatrix} \sqrt{\frac{2}{\tau_{e1}}} \\ \sqrt{\frac{2}{\tau_{e2}}} \end{bmatrix} s_+ \quad (3)$$

in which the subscript '1" and '2' indicate the resonator CCLSR and the VLRR component, respectively, $a_n$ is the mode amplitude of the resonator, $\omega_n$ is the resonance frequency, $1/\tau_{en}$ and $1/\tau_{0n}$ are the decay rates due to radiation and intrinsic losses, respectively, $s_+$ is the excitation signal, and $\sqrt{2/\tau_{en}}$ is the coefficient expressing the degree of coupling between the resonator and the excitation wave. $\omega_1$ is a function of the biasing voltage $V_t$ in the CCLSR and $\omega_2$ is dependent on the oscillation strength of the VLRR. As a result, the overall response of the unit cell (when considering the CCLSR and VLRR components as a whole) is not only modulated by the biasing voltage $V_t$, but is also dependent on the incident RF power strength. The reflection coefficient of the unit cell may be expressed by[31,32]:

$$r = -1 + \frac{\sqrt{\frac{2}{\tau_{e1}}} a_1 + \sqrt{\frac{2}{\tau_{e2}}} a_2}{2|s_+|} \quad (4)$$

The resonance reflection response as a function of incident wave $s_+$ with varying power levels and biasing voltages $V_t$ is plotted in Fig. 1d. Details of the calculations may be found in Supplementary Note 3. As depicted in Fig. 1d, the dips in the curves indicate the resonance frequency and the oscillation strength of the unit cell. At a given biasing voltage (e.g., $V_t = 0$ V), when the excitation power strength is low (e.g., $|s_+| = 0.001$), the unit cell resonates at the designated resonance frequency with a strong oscillation amplitude, which yields a strong magnetic field enhancement in the vicinity of the unit cell. When the excitation power increases, the resonance mode of the unit cell shifts to a lower frequency. In addition, the peak oscillation amplitude at the resonance frequency also decreases along with the frequency shift, which results from the bi-stable nonlinear behavior in the amplitude response of the VLRR[26,31]. The attenuation of the peak amplitude resulting from the strong excitation field ensures that the unit cell will not enhance the magnetic field and not interference with the incident RF wave. Another notable property of the unit cell is the frequency tunability introduced by the controlling circuit. When the biasing voltage is adjusted to 2.5 V or 5 V, as shown in Fig. 1d, the resonance frequency is shifted to a higher value and the variation of the biasing voltage does not impact the self-adaptive response to the incident power. In this fashion, the resonance frequency may be precisely tuned in order to ensure a frequency match without a disruption in the nonlinear resonance response of the metamaterial.

With the aforementioned fundamental mechanism of the metamaterial unit cells, we sought to experimentally validate the resultant advanced properties through the assembly of the meta-atoms in a 4 × 4 array fashion. The unit cells were initially tuned to ~100 MHz, such that the overall resonance frequency of metamaterial approximated 125 MHz due to the coupling between the unit cells. In order to characterize the tunability of the resonance frequency, we employed a coupling loop connected to a network analyzer to excite the metamaterial, using an excitation power of -10 dBm in order to mimic the magnetic reception field $B_1^-$ in MRI. Initially, the metamaterial was configured in a planar shape, with the experimental setup shown in Supplementary Fig. 3a. The reflection coefficients S11 were measured using a sweep of the biasing voltage. From the experimental results plotted in Fig. 1e, as the biasing voltage increases from 0 V to 20 V, the resonance frequency is tuned from 123.5 to 147.0 MHz. Next, in an effort to mirror more optimal geometries for imaging non-planar portions of the human anatomy fashion (wrist, knee, or neck, for example) in a conformal fashion a curved configuration of the metamaterial was analyzed. We measured the reflection coefficients S11 to investigate the tunability of the curved, semi-cylindrical configuration of the metamaterial using an identical experimental setup and sweep parameters; the experimental setup is shown in Supplementary Fig. 3b. As shown in Fig. 1f, a wide resonance frequency tuning range (122.8 ~ 146.5 MHz) is achieved with the adjustment of the biasing voltage from 0 V to 20 V, demonstrating that the curvature of the metamaterial has a negligible effect on its resonance frequency tunability. Importantly, this frequency tuning range is sufficient to compensate for detuning effects during imaging and, thereby, realize an optimized frequency match between the metamaterial and the MRI system. Please see Supplementary Note 4 for further details regarding the degree of metamaterial resonance detuning effect in the presence of varying phantom compositions or metamaterial configured in different shapes. The experimental results plotted in Supplementary Fig. 4 demonstrates the resonance frequency tuning capacity of the metamaterial while considering real-world clinical MRI applications.

Besides the frequency tunability of the metamaterial, the self-adaptive response to the excitation power strength was also verified using an identical experimental setup. Firstly, the metamaterial was configured in a planar shape, with the resonance frequency tuned to ~127 MHz by adjusting the biasing voltage. Next, the reflection coefficients S11, as an indicator of the oscillation amplitude of the metamaterial, were measured with a sweep of the excitation power strength, with the results plotted in Fig. 1g. Initially, the resonance frequency remained constant at a value of ~127 MHz, with minimal change in oscillation amplitude with an excitation power within -8 dBm. As the excitation power strength increased, the resonance frequency shifted to a lower value, with an attenuation of the oscillation amplitude. When the excitation power strength is sufficiently high (e. g. > 7 dBm), an abrupt transition in the spectrum occurred, and the oscillation in the metamaterial was no longer present. In the case of MRI, the RF power during the reception phase is in the μW range, while the power during the transmission phase reaches the kW range. This marked difference in power level between RF reception and transmission phases in MRI is sufficiently large to cause the metamaterial to exhibit its self-adaptive response as a function of excitation power strength[34]. Subsequently, a similar experiment was performed with the metamaterial in its semi-cylindrical configuration, with the test results plotted in Fig. 1h. In this case, an identical response is observed, indicating that the deformation of the metamaterial has no impact on its self-adaptivity. The agreement between the theoretical calculations based on CMT and the experimental results further validates these advanced properties of the metamaterial in its application to MRI.

**2.2 Magnetic field mapping**

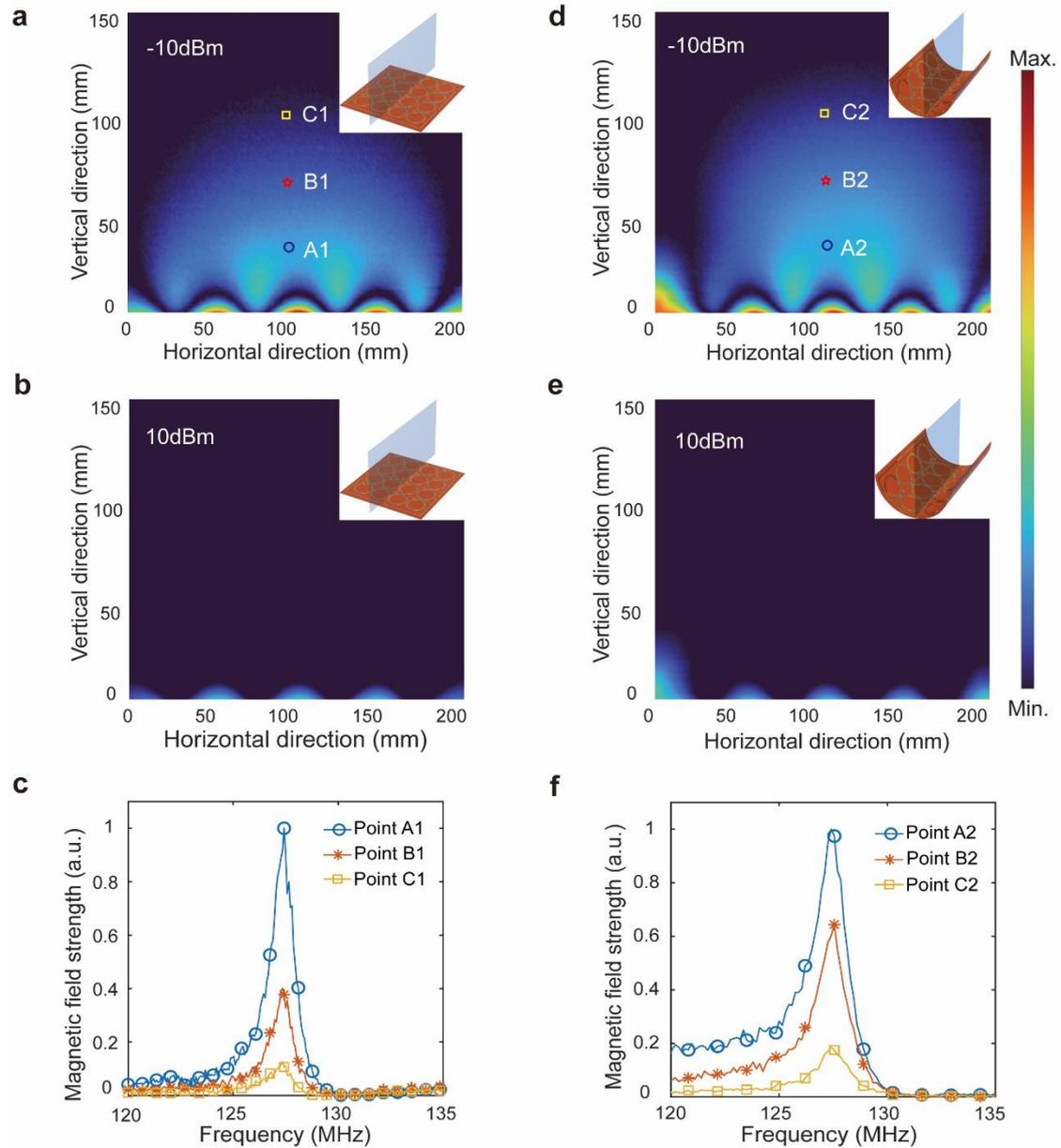

**Fig. 2 | Magnetic field mapping in the vicinity of the metamaterial. a, b,** Experimentally measured magnetic field strength distributed along the planar metamaterial cross-section (depicted as the blue plane in the inset) at different excitation strengths, e.g., $s_+$ = -10 dBm (**a**) and $s_+$ = 10 dBm (**b**). **c,** Spectra of the magnetic field distribution on points A1, B1 and C1, as shown in (**a**). **d, e,** Experimentally measured magnetic field strength distributed along the semi-cylindrical metamaterial cross-section (depicted as the blue plane in the inset) at different excitation strengths, e.g., $s_+$ = -10 dBm (**c**) and $s_+$ = 10 dBm (**d**). **f,** Spectra of the magnetic field distribution on points A2, B2 and C2, as shown in (**d**).

In order to experimentally demonstrate the magnetic field confinement and enhancement, a near field mapping setup was built to measure the magnetic field distribution in the vicinity of the metamaterial. Details of the experimental setup are described in Methods and a schematic of the setup is illustrated in Supplementary Fig. 5. Firstly, the metamaterial in its planar configuration was tuned to ~127 MHz by adjusting the biasing voltage $V_t$, and then excited by an RF power of -10 dBm. The normalized magnetic field distribution on the blue parallel plane above the metamaterial (shown in the inset on Fig. 2a) is depicted in Fig. 2a. For the region (extend from the surface of metamaterial to a distance of ~20 mm), although the magnetic field peaks at the surface of metamaterial, is impractical for application to MRI imaging due to a phase transition between neighboring unit cells. The region (begins at and extends from the surface 20 mm above the metamaterial) is the optimal region for placing a sample to be imaged in MRI. It is evident that the magnetic field peaks close to the center of unit cell, and decays as away from the metamaterial. The field enhancement effect vanishes when ~100 mm away from the metamaterial surface. Using an identical experimental setup, the magnetic field distribution is mapped under a much higher excitation power (10 dBm), with this magnetic field distribution plotted in Fig. 2b. In this case, the magnetic field strength approximates 0, with the metamaterial effectively turned 'off', similar to the absence of the metamaterial. The absence of an appreciable magnetic field in this case again demonstrates the self-adaptive response of the metamaterial to the transmitting field $B_1^+$. In addition, the spectra of the magnetic field strength at three distinct locations 30, 70, and 110 mm away from the metamaterial surface (labeled as points A1, B1, and C1 in Fig. 2a) are plotted in Fig. 2c. The magnetic field is maximal at the resonance frequency regardless of the location measured. With a shift in the resonance frequency, the degree of magnetic field strength is severely compromised. For example, when the frequency shifts by 2 MHz, the field strength decreases by 70%, demonstrating the significance of the frequency matching and the necessity of the capacity for frequency tunability. Next, the magnetic field distribution was mapped for the case in which the metamaterial assumed a semi-cylindrical configuration. The field mapping using excitation powers of -10 dBm and 10 dBm are plotted in Figs. 2d, e, respectively. The field patterns are similar to the case of the planar configuration of the metamaterial, demonstrating that deformation did not impact metamaterial self-adaptive field enhancement. Of importance is the fact that the semi-cylindrical configuration of the metamaterial resulted in deeper penetration depth (~130 mm) of the field enhancement area when compared with the penetration depth of the planar metamaterial (~100 mm), due to the superimposition of the magnetic field enhancement effects on either side of the metamaterial. This increase in penetration depth has significant practical implications in clinical MRI, in which deeper anatomic structures may be of interest. In addition, the spectra of the magnetic field distribution at the same locations are also plotted in Fig. 2f. A similar conclusion may be drawn from the spectra of the semi-cylindrical configuration of the metamaterial: the field enhancement peaks at the resonance frequency, while a frequency mismatch serves to dramatically impair the field enhancement in the vicinity of metamaterial. By comparing the spectra of the magnetic field strength at three points in Figs. 2c, f, one notable advantage of the semi-cylindrical configuration of the metamaterial is that the field strength decays slower than that of the planar metamaterial, yielding a deeper penetration depth of the field enhancement. Besides the experimental results, we also built a numerical simulation model in CST to investigate the magnetic field distribution, including the magnitude and phase information (detailed in Supplementary Note 5). The simulated results are plotted in Supplementary Figs. 6 and 7 for metamaterials in planar and semi-cylindrical configurations, respectively. The experimental near-field mapping results closely align with the numerical simulations of magnetic field strength patterns, affirming the accuracy of our experimental findings. Furthermore, the simulated phase information of the magnetic field in the distant region from the metamaterial indicates that there is no such phase transition region where the original magnetic field and the induced magnetic field cancel each other, thus preventing RF artifacts in the imaging area due to magnetic field cancellation.

## 3. Experimental MRI Validations
## 3.1 MRI validations for metamaterial enhanced body coil

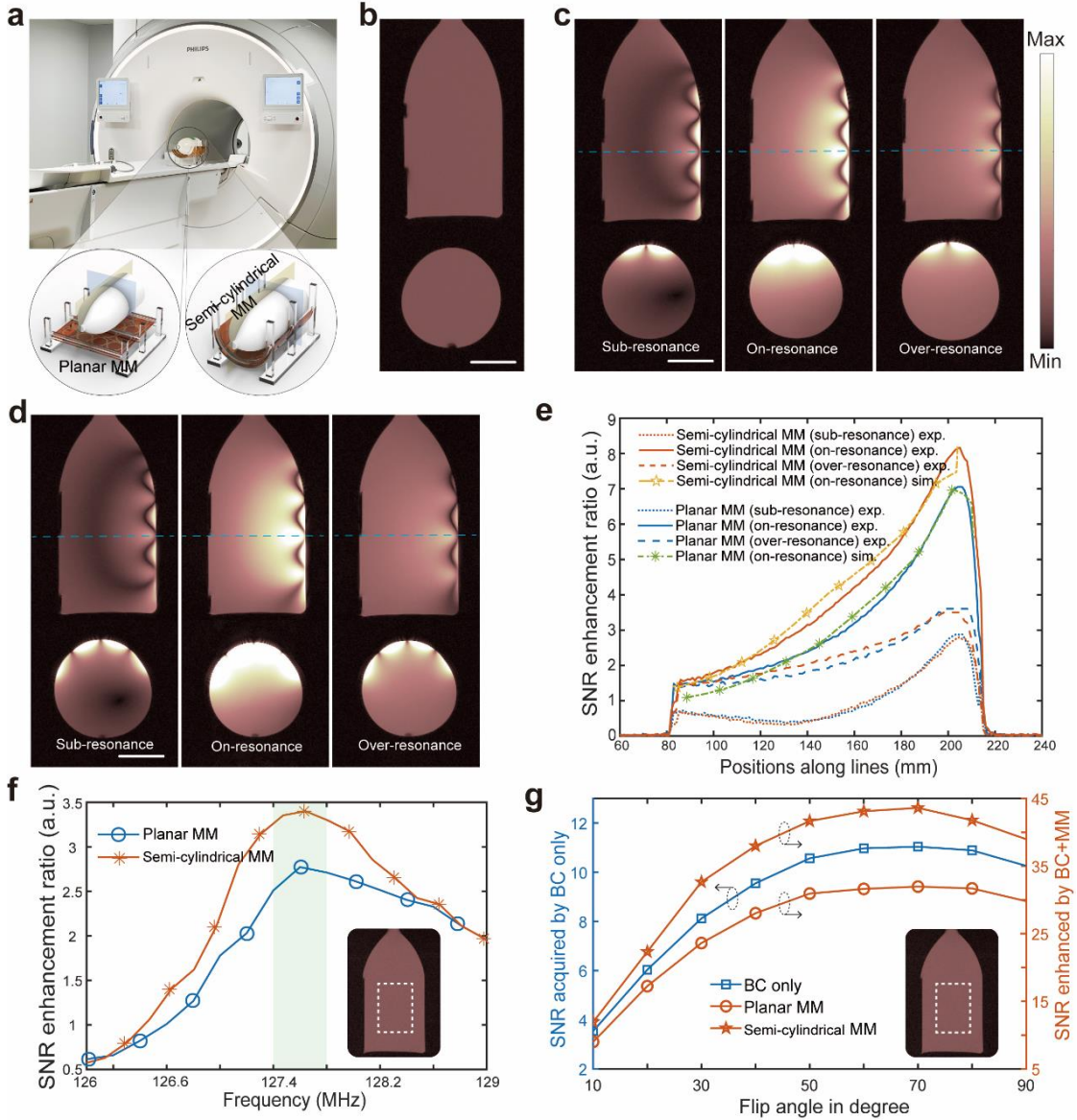

**Fig. 3 | MRI validations for metamaterial enhanced body coil. a,** Experimental setups in the MRI system for metamaterial presenting in planar and semi-cylindrical configurations. **b**, SNR images on sagittal and axial planes captured by BC only, serving as a reference. **c**, SNR images captured by BC enhanced by the planar metamaterial under sub-resonance, on-resonance, and over-resonance conditions. **d**, SNR images captured by BC enhanced by the semi-cylindrical metamaterial under sub-resonance, on-resonance, and over-resonance conditions. **e**, Comparison of the SNR enhancement ratio along the blue dashed lines in (**c**) and (**d**). **f,** SNR enhancement performance as a function of resonance frequency. **g**. Variations of the SNR as a function of flip angle. Scale bars in (**b**), (**c**), and (**d**) are 5 cm.

In order to evaluate the performance of this metamaterial when operating with the body coil (BC) in MRI, experimental validations of the SNR enhancement are performed in a clinical 3T MRI system (Philips Healthcare). The experimental evaluations of SNR were performed using the two-image method (detailed in Methods and Supplementary Fig. 8)[35]. A bottle-shaped phantom filled with mineral oil was initially scanned in the absence of the metamaterial, serving as a reference standard. The SNR reference image of the phantom in the sagittal and axial planes (indicated as the rectangular cutting plane in yellow and the square cutting plane in blue shown in Fig. 3a) are depicted in Fig. 3b. Next, the phantom was placed along the top surface of the metamaterial, with the separation distance between the top surface of the metamaterial and the bottom surface of phantom approximately 5 mm. Subsequently, we fine-tuned the metamaterial's resonance frequency to 126 MHz, thereby leading to a resonance frequency lower than that of the MRI system. The corresponding image, displayed on the left side of Fig. 2c, exhibited noticeable artifacts compared to the reference image obtained without the metamaterial. In this image, there was a region of relative signal loss, and the overall image quality did not show any significant improvement. Following this, we further adjusted the resonance frequency to achieve the optimal condition of resonance matching between the planar metamaterial and the MRI system. The phantom image obtained under the on-resonance condition is featured in the middle of Fig. 2c. Unlike the uniform pattern throughout the reference image, the enhanced areas in the SNR images show similar patterns to the near field mapping results in Figs. 2a, further supporting the direct relevance between the degree of magnetic field enhancement and eventual SNR enhancement during image acquisition. Moreover, the metamaterial-enhanced SNR along the dashed lines in the image are extracted and normalized to the reference, as plotted in Fig. 3e, revealing a substantial enhancement in overall SNR, up to 7-fold when compared to the acquisition without the metamaterial. Next, we further adjusted the resonance frequency of the planar metamaterial to 129 MHz, surpassing the operational frequency of the MRI system. The image acquired under the over-resonance condition is presented on the right side of Fig. 2c. This image does show an SNR gain compared to the reference image; however, the improvement is relatively modest compared to the images obtained with the metamaterial in the on-resonance state. We then replicated the same experimental approach with the semi-cylindrical metamaterial. The images acquired under the sub-resonance, on-resonance, and over-resonance conditions are displayed in Fig. 2d. Similar to the results obtained with the planar metamaterial, artifacts were observed under the sub-resonance condition, while the over-resonance metamaterial showed some SNR improvement, though not as significant as the on-resonance metamaterial. The latter achieved a maximum SNR increase exceeding 8-fold when compared to the reference image acquired without the metamaterial. The SNR enhancement ratio of the semi-cylindrical metamaterial was higher at its maximum and decayed at a slower rate when compared to the planar configuration due to the superimposition effects noted above. To obtain a comprehensive understanding of the correlation between metamaterial enhancement performance and its resonance frequency, we conducted a series of scans of the phantom across a range of resonance frequencies, spanning from 126 to 129 MHz with increments of 0.25 MHz. The resulting SNR images are presented in Supplementary Figs. 9 and 10 for both the planar and semi-cylindrical metamaterials, respectively. Next, we extracted the mean SNR values from the outlined regions (as depicted in Fig. 2f) and presented them as a function of resonance frequency, as illustrated in Fig. 2f. Based on the curve, it can be inferred that the optimal resonance frequency for the metamaterials lies at approximately 127.6 MHz. This finding underscores the significance of frequency tunability for metamaterials when applied in MRI systems, as any frequency mismatch strongly influences SNR enhancement performance. Of note, the optimal frequency for the proposed metamaterials is approximately 127.6 MHz, slightly offset from Larmor frequency in the 3T MRI system, primarily attributed to the phenomenon of over-coupling between the metamaterial and the birdcage coil.

The mechanism of SNR enhancement by the metamaterial was also explored by deriving the governing equation of SNR, which is originally formulated by[25,36]:

$$\text{SNR} \propto \frac{\omega^2 B_c}{\sqrt{R_{BC} + R_{sample} + R_{mm}}} \quad (5)$$

in which $B_c$ is the magnetic field strength generated with unit current in the receiving coil and $\omega$ is the Larmor frequency. As indicated by equation (5), there are three dissipative elements giving rise to the principal noise sources in MRI. The first is the series resistance of the receive coil $R_{BC}$ resulting from the conductive loss of the receive coil, the second is the equivalent series resistor $R_{sample}$ representing the power loss in the sample or patient due to the induced eddy currents, and the third is the equivalent series resistance $R_{mm}$, which represents the conductive power dissipation in the metamaterial. Herein, with the aid of the numerical simulation tool CST, we developed a simple and straightforward method to precisely evaluate the SNR in MRI. Since the Lamer frequency $\omega$ is directly proportional to static field strength $B_0$, this is a constant value for the experimental setup reported herein. As a result, the analytical SNR enhancement ratio could be calculated by the ratio of the simulated magnetic field strength divided by the square root of the total power dissipation in the BC, metamaterial, and phantom, with all of these parameters readily extracted from the simulation results. The analytical results are plotted in Fig. 3e for comparison, showing a high degree of agreement with the MRI experimental results. The SNR simulation model is described in detail in Supplementary Note 6 and Supplementary Fig. 11.

In the case of a conventional metamaterial without self-adaptivity, its introduction would lead to an enhancement of not only RF reception field $B_1^-$ but also transmission field $B_1^+$, yielding marked image artifacts or even imaging failure. As a result, in order to cancel out the enhancement of $B_1^+$ resulting from conventional metamaterials, the transmission energy is generally reduced by decreasing the flip angle (FA) accordingly[17,25]. In order to demonstrate the superiority of our adaptive metamaterial when employed in MRI, another group of MRI validations was performed by imaging the phantom with a sweep in FA from 10º to 90º. The mean value of SNR in the area of interest (outlined by the dashed white rectangle shown in the inset figure of Fig. 3g) as a function of FA is plotted in Fig. 3g. The SNR variations along with the sweep of FAs exhibit the same trend, regardless of the absence or presence of the metamaterial. Furthermore, the SNR value is dramatically boosted in the presence of the metamaterial when compared to the SNR acquired by the BC only. Based on these test results, we may conclude that the metamaterial shows excellent adaptive response to $B_1^+$ and $B_1^-$ in both transmission and reception phases.

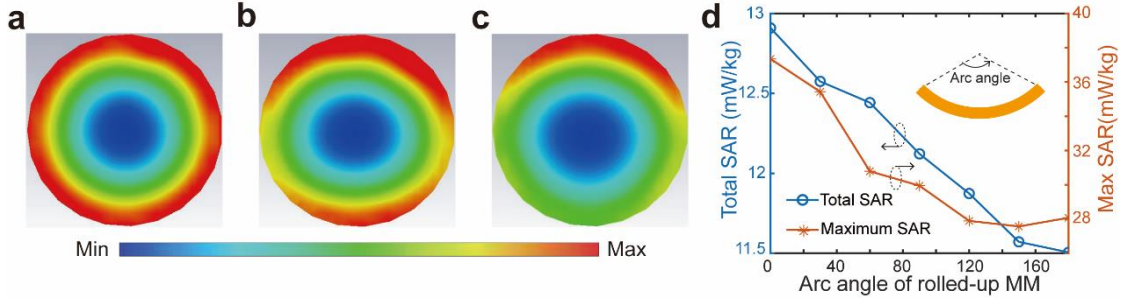

**Fig. 4** | Simulated SAR maps of the phantom in the axial plane with (**a**) BC only, with (**b**) planar metamaterial, and with (**c**) semi-cylindrical metamaterial. **d**, Total and maximum SAR as a function of the arc angle of the rolled-up metamaterials.

In terms of the safety of the metamaterial, SAR is analyzed with and without the presence of the metamaterial. SAR refers to the instantaneous rate of energy dissipated by a RF field in a certain mass of tissue, given by[37,38]:

$$SAR = \frac{\sigma E_p^2}{2\rho} \qquad (6)$$

in which σ and ρ are the conductivity and mass density of the medium, respectively. $E_P$ is the peak amplitude of the time-varying electric field, which is induced by the transmitted RF irradiation in MRI. Obtaining an accurate SAR estimation is generally complicated given a sample having complex geometry and nonuniform distribution in its electrical properties, especially given an irregular distribution of the RF field with the introduction of a metamaterial[39]. Aided by CST (the simulation model is similar to SNR simulation, as described in Supplementary Note 6 and Supplementary Fig. 11), the SAR maps (normalized to 1W accepted power) in the axial plane were simulated in the absence of the metamaterial as reference and in the presence of planar and semi-cylindrical configurations of the metamaterial, respectively, as depicted in Figs. 4a-c (all the SAR maps are normalized to the same scale bar). SAR tends to be concentrated about the periphery, as indicated in the SAR maps. Through the comparisons of the three maps, the introduction of the metamaterial does not increase the SAR due to the self-adaptive detuning resonant response to the field $B_1^+$. Instead, the SAR in the area surrounding the metamaterial is clearly suppressed, due to the fact that the electric field is highly confined around the narrow capacitive gap between neighboring conducting wires in the metamaterial. Besides the SAR maps, the maximum and total SAR levels with 1g averaging mass are extracted and plotted in Fig. 4d for a quantitative comparison. Both the maximum and total SAR is reduced in the presence of the metamaterial, with the semi-cylindrical configuration of the metamaterial exhibiting a superior performance when compared with the planar metamaterial due to its relatively large coverage area of the phantom. The electric field in BC, as well the SAR patterns when the metamaterial is rolled-up at various arc angles, are described in detail in Methods, Supplementary Note 7 and Supplementary Figs. 12 and 13.

**3.2 MRI validations for metamaterial enhanced surface coil**

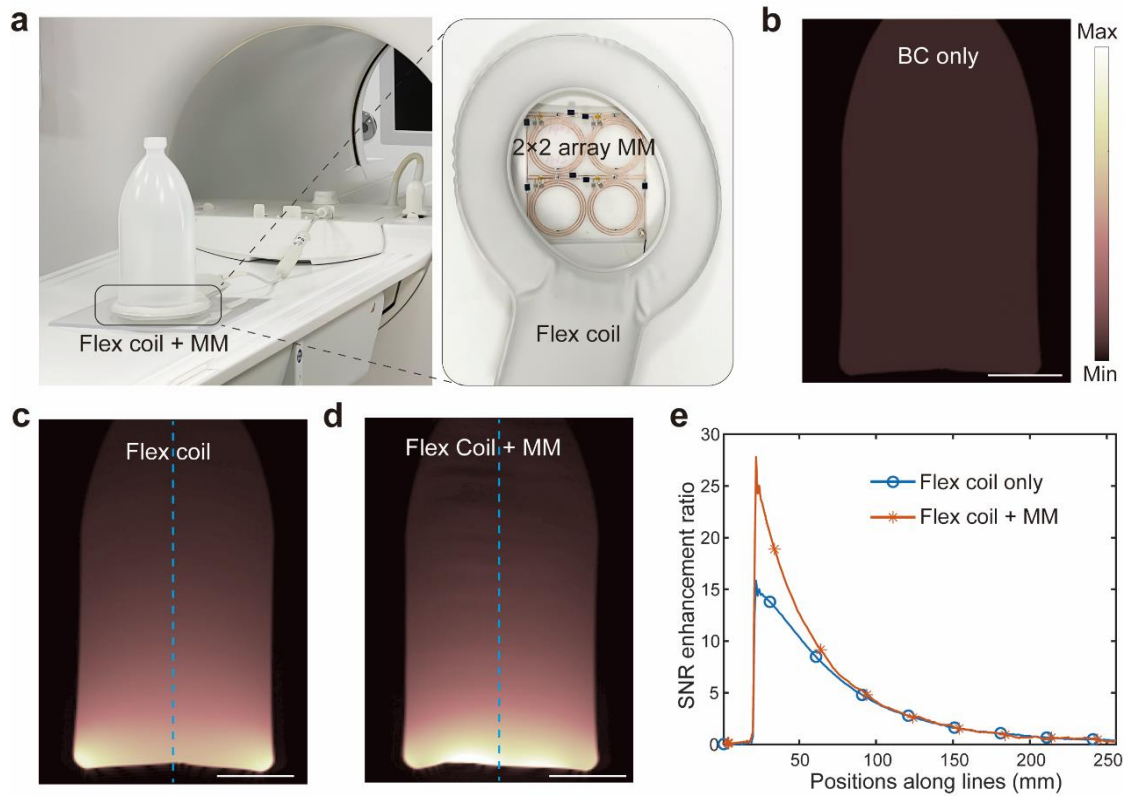

**Fig. 5 | MRI validations for the surface coil enhanced by metamaterial. a,** Experimental setups in the MRI system. Inset: the configuration of surface coil integrated with metamaterial. **b**, SNR images captured by the BC only. **c**, SNR images captured by the surface coil only serving as a reference. **d**, SNR image captured by the surface coil combined with the metamaterial. **e**, Comparison of the SNR enhancement ratio along blue dashed lines in (**c**) and (**d**). Scale bars in (**b**), (**c**), and (**d**) are 5 cm.

In this work, our primary objective was to design an advanced metamaterial to enhance the SNR of MRI as an additive technology. This metamaterial is not intended to replace current commercial surface coils but rather to augment the imaging capabilities of existing receive coils, including both body coil and surface coils, through their integration. Thus, we have demonstrated that the metamaterial may be combined with the BC to boost the imaging power of MRI. More importantly, the metamaterial presented here has the potential to be integrated with surface coils to boost their imaging power. To investigate the performance when combining the metamaterial with surface coils, we constructed a metamaterial in a 2 × 2 array fashion and integrated them into the dStream Flex M coil (Philips Healthcare). This integrated surface coil and the experimental setup during scanning are depicted in Fig. 5a. Initially, the phantom was scanned separately using the BC and Flex coil only, without the integration of the metamaterial. These scans served as reference images, and the corresponding results are shown in Figs. 5b and c. Subsequently, employing identical scanning parameters and imaging sequences, we conducted another scan using the Flex coil with the metamaterial integrated. The resulting SNR image is presented in Fig. 5d. For quantitative comparisons, we extracted SNR values along dashed lines in Figs. 5c and d and normalized them to the BC only reference image, as shown in Fig. 5e. Comparing the uniform SNR pattern observed in the BC only reference image, the switch from the BC to the Flex coil resulted in a significant and dramatic improvement in SNR. Specifically, there was an approximately 15-fold increase in SNR at the bottom of the phantom near the Flex coil. More importantly, the integration of the metamaterial into the Flex coil led to a nearly 2-fold increase in SNR compared to the Flex coil only image. These MRI validations demonstrate that metamaterials have the potential to not only enhance the imaging capabilities of the BC but also significantly improve the performance of surface coils. This opens up new possibilities for the application of metamaterials in MRI across a wide range of scenarios.

## 3.3 MRI validation with ex vivo samples

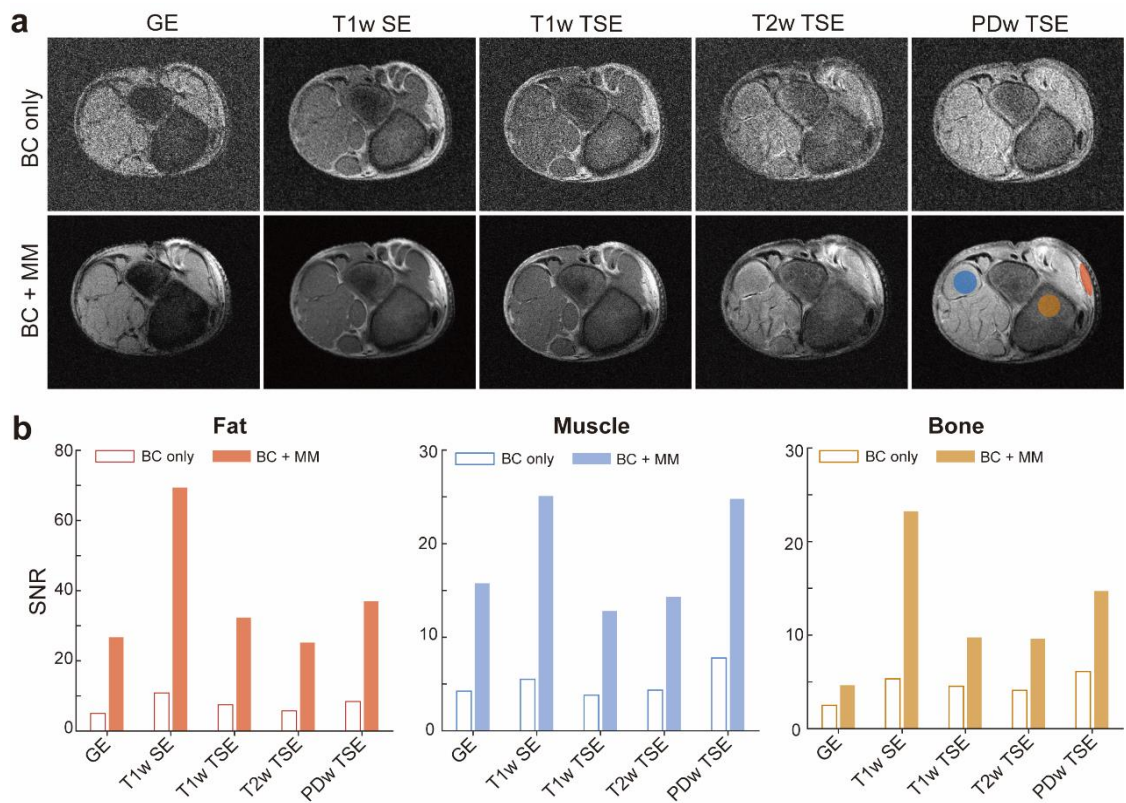

**Fig. 6 | MRI scans of ex vivo porcine leg by BC through different imaging sequences. a,** SNR images captured with and without the presence of metamaterial. **b,** Quantitative assessment of the SNR performance of specific tissues.

Besides the mineral oil phantom, an ex vivo sample of porcine leg was also employed in the experimental MRI validations with BC to preliminarily demonstrate the performance of the metamaterial in biomedically-relevant imaging. As opposed to the gradient echo (GE) imaging employed for the mineral oil phantom, T1 weighted turbo spin echo (T1w TSE) imaging featuring a series of 180º-refocusing pulses after a single 90º excitations pulse, was employed for this MRI validation. The porcine leg was wrapped by the flexible metamaterial, and 8 slices (images) on the axial plane are scanned in the absence and presence of the metamaterial, the images are depicted in Supplementary Fig. 14. These images provide a preliminary demonstration of the potential of our technology to enhance SNR in complex anatomical settings. From the multiple slices obtained with T1w TSE, we picked the cutting plane of image slice #5 and conducted scans of the porcine leg on the same plane using another four mainstay pulse sequences commonly used in clinical MRI: GE,

T1-weighted spin echo (T1w SE), T2-weighted turbo spin echo (T2w TSE), and proton density-weighted turbo spin echo (PDw TSE). The slices (images) scanned in the absence and presence of metamaterials are depicted in Fig. 6a. Notably, when compared to images acquired by using the BC only, images with enhanced SNR were achieved in the presence of the metamaterial even though different tissues are included in the leg (fat, muscle, bones, and bone marrow). Additionally, the result proves that a metamaterial featuring a self-adaptive property readily operates with a variety of clinical RF transmission pulse sequence frequently used in MRI. For quantitative analysis of SNR enhancement in specific tissues, we plotted bar graphs of SNR mean values for muscle, fat, and bone, outlined by dashed lines in Fig. 6b. The SNR values and corresponding enhancement ratios reveal a substantial 2~5 fold increase in SNR attributed to the introduction of the metamaterial. Additionally, a larger biological sample, a pineapple, was scanned with the metamaterial in an arc shape between the planar and semi-cylindrical shape, with the scan images depicted in Supplementary Note 8 and Supplementary Fig. 15, further highlighting the potential utility of metamaterial-enhanced MRI.

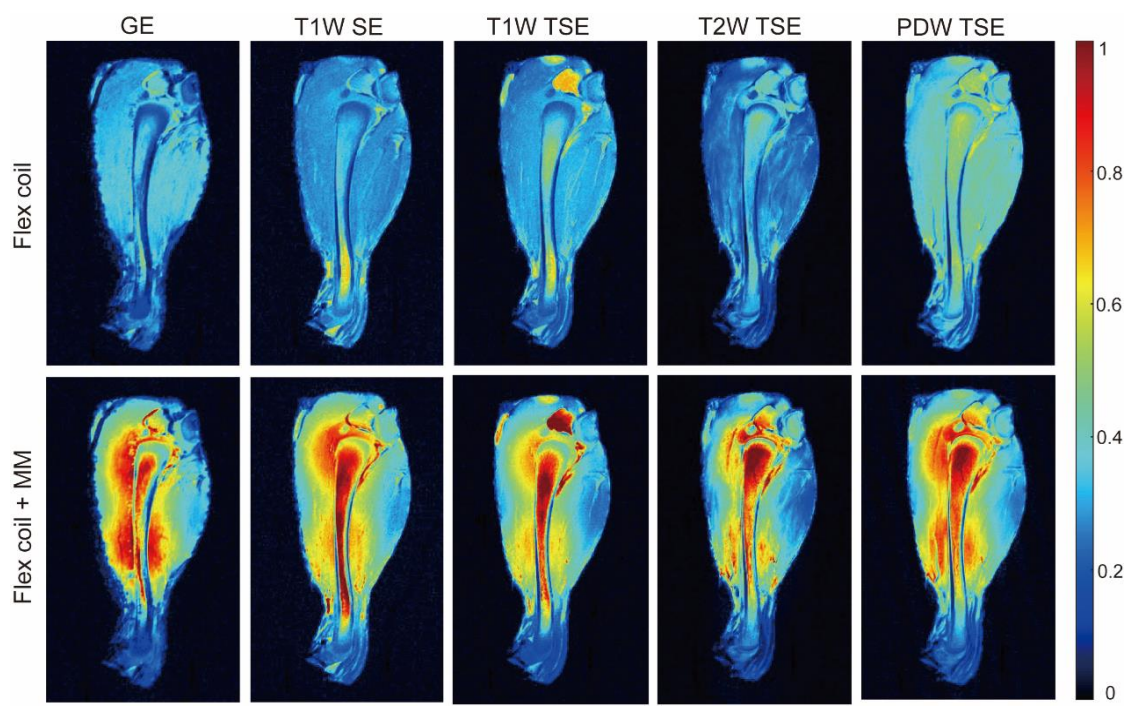

**Fig. 7 | Comparisons between images captured by Flex coil in the absence and presence of metamaterial.**

Finally, we sought to validate the performance of surface coil enhanced by the metamaterial with an ex vivo chicken leg. Using the setup shown in Fig. 5a, we placed the chicken leg on top of the Flex coil and employed the same scanning sequences previously used with the porcine leg. SNR maps of the chicken leg were generated by normalizing to the background noise level, as shown in Fig. 7. By comparing the images acquired using the Flex coil with and without the integrated metamaterial, we observed that the metamaterial has the potential to enhance not only the performance of the BC but also that of surface coils. This suggests its applicability for improving the imaging capabilities of MRI across various clinical RF transmission pulse sequences commonly used in MRI.

## Conclusion

This work demonstrates a conformal, intelligent metamaterial, capable of selectively boosting the $B_1^-$ field under an optimized frequency matched condition and switching to an off-resonance self-adaptively without interference with the $B_1^+$ field, yielding marked improvements in SNR enhancement performance when employed in MRI. In contrast to previous conventional metamaterials applied to MRI, the meta-atom of the reported metamaterial is designed by integrating a tunable CCLSR and a nonlinear VLRR, yielding the properties of voltage-driven frequency tunability and an RF field dependent resonance response. Attributed to these advanced features, the metamaterial has the potential to serve as an auxiliary device, enhancing not only the imaging capabilities of the body coil but also those of surface coils. Consequently, metamaterials exhibit significant promise for widespread utilization within MRI systems. The mathematical modeling based on CMT and the following on-bench characterizations of the metamaterial elucidate its governing physical mechanism, providing a profound analysis to gain a deeper understanding of the frequency-tunable and field dependent resonant behavior of artificial materials. The metamaterial reported herein is applicable to different 3T MRI systems, regardless of the manufacturers. Furthermore, the metamaterial is readily adjusted to be applicable to MRI system with different static magnetic field strengths, such as 1.5, 7.0, or 9.4 T. Since these superior properties of metamaterials are predominantly derived from their constituent meta-atoms, the metamaterial properties theoretically remain valid with different numbers of unit cells. Thus, the metamaterial is not limited to 4×4 unit cell configurations and may be readily tailored into different shapes consisting of arbitrary numbers of unit cells for specific application scenarios. Ultimately, the voltage-driven frequency tuning mechanism and field-dependent resonance response offer a promising pathway for the future development of application-oriented electromagnetic devices beyond the application to MRI.

## Methods

**Geometry and fabrication of metamaterial.** The reported metamaterial was fabricated using 0.089 mm thick copper foil tape, which was supported on a flexible dielectric layer polyimide film (Kapton®) having dimensions of 305 × 305 mm and a thickness of 0.013 mm. The CCLSR had an inner radius of 20.4 mm, a strip width of 0.8 mm, and an inter-strip spacing of 0.8 mm. The VLRR, sharing the same concentric point with the CCLSR, had an inner radius of 18.8 mm and a strip width of 0.8 mm. The metamaterial was realized by assembling a 4 × 4 array of the unit cells, in which the column separation distance of the neighboring unit cells is 45.6 mm, and the row separation distance is 48.8 mm. The biasing voltage lines were inserted between the neighboring rows for applying biasing voltage to the controlling circuits of each unit cell.

**Characterization of the frequency tunability of the metamaterial.** We employed a vector network analyzer (VNA, E5071C, Keysight Inc) with an inductive loop to excite the magnetic resonance of the metamaterials, in which the excitation power was set to -10 dBm to mimic the $B_1^-$ field in MRI. The reflection spectra were measured with a sweep in biasing voltage from 0 to 20 V with a step of 0.5 V. In the reflection spectra S11, the dips correspond to the resonance mode of the metamaterials. In the characterization, the metamaterial was measured in both planar and semi-cylindrical configurations.

**Characterization of the self-adaptivity to excitation strength of the metamaterial.** Similar to the tunability characterization, the VNA with an inductive loop was employed to excite the magnetic resonance of the metamaterials. In the characterization, the biasing voltage was adjusted to tune the resonance frequency of the metamaterial to ~127 MHz. The reflection spectra S11 were measured with a sweep in excitation power from -30 to 10 dBm with a step of 1 dBm.

**Magnetic field mapping.** In this setup, a coupling loop 250 mm in diameter was connected to port 1 of the VNA. RF signal fed into the coupling loop served as the excitation wave. A small circular loop with a diameter of 10 mm served as the probe which was connected to port 2 of the VNA via a coaxial cable. The judiciously designed probe size was small enough to reduce its impact on the field distribution of the metamaterial, ensuring a high spatial resolution of the magnetic field mapping, while also being large enough to sense the magnetic field near the metamaterial when excited by a low excitation power. The probe was mounted to a two-dimensional motorized stage which was controlled by a computer, enabling data collection at a widely diverse distributions of different locations surrounding the metamaterial. 200 × 150 data points were collected on cutting planes (indicated in Figs. 3a-d) of 200 × 150 mm in area. The transmission coefficient (i.e., S21) was measured to indicate the magnetic field strength.

**MRI validations with phantom.** A gradient echo(GE) imaging sequence was employed using a repetition time (TR) and echo time (TE) of 100 ms and 4.6 ms, respectively. GE imaging was first performed to capture a phantom image (as shown in Supplementary Fig. 8a), followed by capturing a noise image by shutting down the transmission RF coil (as shown in Supplementary Fig. 8b). The SNR images of the phantom were calculated by the ratio between the mean value of magnitude phantom image and the standard deviation of the noise image.

**MRI validation with ex vivo samples.** The ex vivo samples of the porcine leg and chicken leg were employed to mimic lower limb anatomy imaging, demonstrating the performance of the metamaterial in biomedically-relevant imaging. The porcine leg samples used in this work were obtained from a local butcher shop.

**Numerical simulation.** The numerical simulations were performed with CST Microwave Studio software. In the simulation model, the dimensions of the metamaterial were the same as the fabricated sample described above. The mineral oil phantom was modeled by a material with relative permittivity of 2.1, electric conductivity of 0.175 S/m, and material density of 800 kg/m$^3$.

**Data availability**
The data that support the findings of this study are available from the corresponding author upon reasonable request.

**Acknowledgements**
This research was supported by the National Institute of Health (NIH) of Biomedical Imaging and Bioengineering Grant No. 1R21EB024673. The authors also thank the Rajen Kilachand Fund for Integrated Life Science and Engineering and Boston University Photonics Center for technical support. The authors are grateful to Dr. Yansong Zhao for his experimental assistance during the MRI testing.

**Author Contributions**

X. Zhang, S. W. Anderson, and K. Wu conceived the study. K. Wu, X. Zhao, and X. Zhang conducted the numerical modeling and theoretical analysis, K. Wu, X. Zhu, and X. Zhang constructed the metamaterials. K. Wu, X. Zhu, S. W. Anderson, and X. Zhang designed and conducted the experiments. All authors participated in discussing the results. X. Zhang, S. W. Anderson, and K. Wu wrote the manuscript.

**Conflict of Interest**

The authors have filed a patent application on the work described herein, application No.: 16/002,458 and 16/443,126. Applicant: Trustees of Boston University. Inventors: Xin Zhang, Stephan Anderson, Guangwu Duan, and Xiaoguang Zhao. Status: Active.

**Additional information**

Supplementary information is available for this paper.